\documentclass[aps,prb,reprint,twocolumn,showpacs,floatfix,superscriptaddress]{revtex4}
\usepackage{graphicx}
\usepackage{amssymb}
\usepackage{amsmath}
\usepackage{lmodern}
\usepackage{color}
\begin{document}

\preprint{This line only printed with preprint option}

\title{Localized magnetic moments in a Dirac semimetal as a spin model with long--range interactions}

\author{E. Kogan and M. Kaveh}

\affiliation{Jack and Pearl Resnick Institute, Department of Physics, Bar-Ilan University, Ramat-Gan 52900, Israel}
\affiliation{Cavendish Laboratory, University of Cambridge, J J Thomson Avenue, Cambridge CB3 0HE, UK}

\date{\today}

\begin{abstract}
We   connect  between the problem of thermodynamics of  localized magnetic moments in a Dirac semimetal,
the interaction with  relativistic electrons leading to the
effective ferromagnetic exchange  between the moments,
and the  existing theories dealing with  long--range exchange interaction.
We point out that the results of high--temperature expansion for the free energy of a dilute ensemble of magnetic impurities in the semimetal
performed by V. Cheianov et al. (Phys. Rev. B {\bf 86}, 054424 (2012)) give an indication to the existence of a new disordered fixed point
 in such model.
\end{abstract}


\maketitle

Long-distance exchange interaction between magnetic moments mediated by the mobile
carriers is known as the Ruderman-Kittel-Kasuya-Yosida
(RKKY) exchange \cite{ruderman}. Recently there appeared interest in RKKY interaction  in the  class of  materials in which the low-energy electron excitations
resemble massless Dirac particles: graphene \cite{novoselov,castro}, chiral metals
formed at the surface of topological insulators \cite{hsieh,hasan}, and silicene \cite{aufray}.
There is a peculiarity of the RKKY exchange in such
conductors  which make
them qualitatively different from usual metals:  the Friedel oscillations
are either absent or commensurate with the lattice \cite{brey}.
In particular,
quite a few papers studied  collective behavior of magnetic adatoms randomly distributed on the surface of a topological
insulator \cite{liu,abanin,rosenberg,chulkov}.

Our communication is inspired by the very interesting publication by Cheianov et al. \cite{cheianov}.
The authors considered the high--temperature expansion in the inverse temperature for the disorder averaged  magnetic susceptibility of a dilute ensemble of Ising magnetic impurities in a 2d Dirac
 semimetal.  From this expansion they found the critical temperature of the ferromagnetic phase transition
and the magnetic susceptibility critical exponent.

 We suggest that high--temperature
expansions
can shed light on  critical behavior of the long--range exchange
interaction models, both with and without quenched disorder. That is why we decided to briefly sum up
vast body of existing results in the field, obtained by the renormalization group (RG) analysis, numerical simulations, etc,
emphasising still open problems in the theory.

RKKY effective exchange interaction between a pair of localized magnetic moments described by spins ${\bf S}_1$ and ${\bf S}_2$ has a very simple structure
\begin{eqnarray}
H_{RKKY}=-J(R){\bf S}_1{\bf\cdot  S}_2,
\end{eqnarray}
where $J(R)=I^2\chi(R)$,
$I$ is the exchange interaction between the localized magnetic moment and  itinerant electrons, $R$ is the distance between the magnetic moments, and
\begin{eqnarray}
\label{abr}
\chi(R)=-\frac{1}{4}\int_{0}^{1/T}{\cal G}({\bf R};\tau){\cal G}(-{\bf R};-\tau)d\tau
\end{eqnarray}
is the free electrons static real space spin susceptibility \cite{kogan}. The Matsubara Green's function ${\cal G}$ is \cite{abrikosov}
\begin{eqnarray}
{\cal G}({\bf R},\tau)=-\left\langle T_{\tau} c({\bf R},\tau)c^{\dagger}({\bf 0},0) \right\rangle.
\end{eqnarray}

Further on we assume that $T=0$ and the  Fermi energy is at the Dirac points. Then   the Green's function is
\begin{eqnarray}
\label{rk}
{\cal G}(R;\tau)=-\text{sign}(\tau)\Omega\int \frac{d^d{\bf k}}{(2\pi)^d}
e^{i{\bf k \cdot R}-vk|\tau|},
\end{eqnarray}
where $d$ is the dimensionality of the space,   $\Omega$ is the volume of the elementary  cell, and $v$ is the velocity of electrons.
Performing  integration in Eq. (\ref{rk})
we get
\begin{eqnarray}
{\cal G}(R;\tau)\sim\frac{\text{sign}(\tau)\Omega v|\tau|}{\left(R^2+v^2\tau^2\right)^{3/2}},\;\; d=2\\
{\cal G}({\bf R};\tau)\sim \frac{\text{sign}(\tau)\Omega Rv|\tau|}{\left(R^2+v^2\tau^2\right)^2},\;\; d=3.
\end{eqnarray}
Next performing  integration in Eq. \eqref{abr} we obtain
\begin{eqnarray}
\label{2d}
\chi\left(R\right)\sim\frac{\Omega^2}{ vR^3},\;\; d=2\\
\label{3d}
\chi\left(R\right)\sim\frac{\Omega^2}{vR^5},\;\; d=3.
\end{eqnarray}
(Actually Eqs. \eqref{2d} and \eqref{3d} can be obtained just from dimensionality considerations; numerical coefficients are anyhow of no interest to us.)

The problem of thermodynamics of magnetic moments, forming a periodic lattice, with  an isotropic $n-$component order parameter and 
 algebraically decaying ferromagnetic exchange interactions $J(R)$ 
\begin{eqnarray}
\label{long}
J(R)\sim 1/R^{d+\sigma},
\end{eqnarray}
 corresponds to the effective $O(n)$ Hamiltonian \cite{sak}
\begin{eqnarray}
\label{mi}
H=\int d^dx\left[\frac{b}{2}\left(\nabla^{\sigma/2}\vec{\phi}\right)^2
+\frac{c}{2}\left(\nabla\vec{\phi}\right)^2+\frac{r}{2}\vec{\phi}^2+\frac{g}{8n}\left(\vec{\phi}^2\right)^2\right].\nonumber\\
\end{eqnarray}
(Transition from the discreet exchange Hamiltonian to Landau-Ginzburg one is  explained, for example, in Ref. \cite{vaks}.)
For the short--range exchange $b=0$. For the long--range exchange $b,c\neq 0$; the term  $\left(\nabla\vec{\phi}\right)^2$ is   generated dynamically even if the initial microscopic Hamiltonian is purely long--range.

Study of such a Hamiltonian has a very long history.
We refer the reader to Ref. \cite{luijten} for the list of works published before 1997.
The phase diagram for this model was proposed
in an early  seminal contribution  \cite{fisher}.  For $d>\text{min}(2\sigma,4)$
the model is characterised by  Gaussian  fixed point. In this
regime for all $n$ we have
\begin{eqnarray}
\label{si}
\eta_G &=& 2-\sigma \\
\gamma_G &=& 1\\
\label{si2}
\nu_G &=& 1/\sigma.
\end{eqnarray}
Eqs. \eqref{si} and \eqref{si2}  differ from  those of Landau-Ginzburg theory: $\eta=0$, $\nu=1/2$.
This
is due to the fact that for the Hamiltonian \eqref{mi} field correlation function
in paramagnetic phase (in momentum representation and for small momentum)
in the mean field approximation is
\begin{eqnarray}
\label{gl}
G(q)=\frac{1}{bq^{\sigma}+r},
\end{eqnarray}
in distinction to the traditional one
\begin{eqnarray}
\label{gs}
G(q)=\frac{1}{cq^2+r}.
\end{eqnarray}
(In both cases $r\sim t\equiv (T-T_c)/T_c$.) The  critical exponent $\eta$ is defined \cite{ma} by Equation (for $T=T_c$)
\begin{eqnarray}
G(q)\sim q^{-2+\eta},
\end{eqnarray}
which explains Eq. \eqref{si}.

For $d<\text{min}(2\sigma,4)$ two non Gaussian fixed points
compete with each other: the Wilson--Fisher or short--range (SR) fixed point \cite{wilson}, meaning that the model is equivalent to one with
short--range interactions and the Fisher--Ma--Nickel or long--range (LR) fixed point \cite{fisher}, specific for long--range interaction.
The case \eqref{2d}  lies on the boundary of the "classical" region and the long--range fixed point region,
where the observables
 differ from those of the mean field theory by logarithmic factors.
In the critical region above the critical temperature $T_c$ the correlation length. susceptibility and heat capacity   vary as
 \cite{fisher,larkin}
\begin{eqnarray}
\label{fisher}
&&\xi(T)\sim t^{-1}\left(\ln t^{-1}\right)^{n'}\nonumber\\
&& \chi(T)\sim t^{-1}\left(\ln t^{-1}\right)^{n'},\\
&& C\sim\left(\ln t^{-1}\right)^{(4-n)/(n+8)} (n<4);\;\; C\sim\ln \ln t^{-1} (n=4).\nonumber
\end{eqnarray}
where  $n'=(n+2)/(n+8)$.
Thus for Ising model $n'=1/3$, for $XY$ model $n'=2/5$, and for isotropic Heisenberg model $n'=5/11$.
For $n>4$ the specific heat remains finite and has no jump.
In the critical region below the critical temperature $T_c$ the spontaneous magnetization  varies as
\begin{eqnarray}
m(T)\sim t^{1/2}(\ln t^{-1})^{6/(n+8)}.
\end{eqnarray}
It is worth mentioning that the predicted logarithmic corrections were accurately observed in extensive Monte Carlo simulations of Ising models
\cite{luijten}. It would be interesting to see to what extent do these results correspond to one obtained from high--temperature expansions?.
We wonder, whether a kind of special treatment of hig--temperature expansions proposed in Ref. \cite{hellmund} to extract logarithmic corrections
can be of some help.

Here probably a simple explanation, why the  upper critical dimension in the model with short--range exchange is 4, and
in the model with long--range exchange \eqref{long} is $2\sigma$ (for  $\sigma<2$), would be relevant.  Scaling transformation of the Hamiltonian \eqref{mi} starts with writing down Hamiltonian \eqref{mi} in momentum representation (integration  with respect to $q$   is limited by some ultraviolet cutoff $\Lambda$). We perform integration with respect  to $q$, satisfying $\Lambda/s<q<\Lambda$, where $s\gg 1$ \cite{patashinskii}. In second order of perturbation theory in $g_0$ the only graph which is necessary to take into account is proportional to
\begin{eqnarray}
\label{symbol}
\int_{\Lambda/s<q<\Lambda} \frac{d^dq}{(2\pi)^d}G^2(q),
\end{eqnarray}
where Green's function is calculated for $a=0$.
Integral \eqref{symbol} with Green's function \eqref{gs}  is
\begin{eqnarray}
\int_{\Lambda/s<q<\Lambda} \frac{d^dq}{(2\pi)^d}G^2(q)\sim \int_{\Lambda/s}^{\Lambda}dqq^{d-5}\sim\frac{\Lambda^{-\epsilon}\left(s^{\epsilon}-1\right)}{\epsilon},\nonumber\\
\end{eqnarray}
where $\epsilon=4-d$.
On the other hand, for Green's fuction\eqref{gl}
\begin{eqnarray}
\int_{\Lambda/s<q<\Lambda} \frac{d^dq}{(2\pi)^d}G^2(q)\sim \int_{\Lambda/s}^{\Lambda}dqq^{d-1-2\sigma}\sim\frac{\Lambda^{-\epsilon}\left(s^{\epsilon}-1\right)}{\epsilon},\nonumber\\
\end{eqnarray}
where $\epsilon=2\sigma-d$. In both cases logarithmic dependence upon $s$
\begin{eqnarray}
\int_{\Lambda/s<q<\Lambda} \frac{d^dq}{(2\pi)^d}G^2(q)\sim \ln s
\end{eqnarray}
corresponds to $\epsilon=0$.

In the lowest order  approximation (with respect to $g$),
of two terms   $bq^{\sigma}$ and $cq^2$ the term with the lower degree is relevant and the term
with the higher degree is irrelevant.
Thus the transition between the fixed points for $d=4$ corresponds to $\sigma=2$.
Fisher et al. \cite{fisher} assumed that this remains true for any $d$, and,  while in the  long--range fixed point region the exponent $\gamma$ is a nontrivial function of $\sigma$ and $d$,  simple Eq. \eqref{si}
is valid  there for the exponent $\eta$.

The last statement can be justified, in particular, in the large--$n$ limit of $O(n)$ model \cite{zin},
as it was done in Ref. \cite{bresin}.
We put in Hamiltonian \eqref{mi}  $c=0,b=1$.
and introduce the auxiliary imaginary field $\lambda(x)$ conjugate to $\vec{\phi}^2$:
\begin{eqnarray}
\exp\left(-\int d^dx\left[\frac{r}{2}\vec{\phi}^2+\frac{g}{8n}\left(\vec{\phi}^2\right)^2\right]\right)\nonumber\\
\sim\int D\lambda\;\exp\left(\int d^dx\left[\frac{n}{2g}\lambda^2-\frac{nr\lambda}{g}-\frac{\lambda}{2}\vec{\phi}^2\right]\right).
\end{eqnarray}
We keep the longitudinal component  $\phi_1$ (fixed
by a vanishing external field)  and integrate on $n-1$ transverse components of $\vec{\phi}$. We finally set $\phi_1 =\sqrt{n}\varphi$ and
arrive to a reduced
Hamiltonian:
\begin{eqnarray}
H(\lambda,\varphi)&=&n\int d^dx\left[\frac{1}{2}\left(\nabla^{\sigma/2}\varphi\right)^2
+\frac{\lambda}{2}\varphi^2+\frac{r\lambda}{g}-\frac{\lambda^2}{2g}\right]\nonumber\\
&+&\frac{n-1}{2}\text{Tr}\ln\left[-\Delta^{\sigma}+\lambda\right].
\end{eqnarray}
The large $n$ limit is thus given by the saddle point equations in the two fields $\varphi$ and $\lambda$ and
the corrections are the usual loop expansion. In the absence of space varying external field,
we obtain the equations
\begin{eqnarray}
\varphi\lambda&=&0\nonumber\\
\lambda-r-\frac{g}{2}\varphi^2&=&\frac{g}{2}\int\frac{d^dq}{(2\phi)^d}\frac{1}{q^{\sigma}+\lambda}.
\end{eqnarray}
The previous equations can be solved easily.
At and above $T_c$ the magnetization $\varphi$ vanishes, at $T_c$ $\lambda= 0$, above $\lambda\neq 0$. Thus above $T_c$ the saddle-point equation reads
\begin{eqnarray}
\label{25}
\frac{t}{\lambda}=\frac{2}{g}+\frac{1}{(2\pi)^d}\int\frac{d^dq}{q^{\sigma}\left(q^{\sigma}+\lambda\right)},
\end{eqnarray}
where
$r-r_c=\frac{g}{2}t$,
and $t$ is proportional to $T-T_c$.
For $d > 2\sigma$ the integral converges when $\lambda$ vanishes and one obtains the mean field result
\begin{eqnarray}
\xi=\lambda^{-1/\sigma}\sim t^{-1/\sigma},
\end{eqnarray}
i.e. $\nu=1/\sigma$.
For $\sigma < d < 2\sigma$ the integral in the  r.h.s. of \eqref{25} diverges near $T_c$ as $\lambda^{d/\sigma-2}$ i.e.
\begin{eqnarray}
\nu=1/(d-\sigma),
\end{eqnarray}
and from the scaling law $\nu(d-2+ \eta) = 2\beta$ and the relation $\beta=1/2$ valid in given approximation one recovers Eq. \eqref{si2}.

Although the general outline of the phase diagram \cite{fisher}  has been
widely accepted, the location of the boundary between the SR and the LR fixed point has become the
scene of a debate. The reason of the objections to the initial position of such boundary at $\sigma=2$ is simple.
From the conjecture  follows that $\lim_{\sigma\to 2}\eta=0$.
Together with this for $\sigma>2,d<4$ the critical exponents assume their SR values, with positive value of $\eta_{SR}$.
Then it would imply a jump
of the exponent $\eta$ from 0 up to $\eta_{SR}$ at $\sigma = 2$. This contradiction
was  removed by Sak \cite{sak}, who, by taking into account
higher order terms in the RG calculations, predicted that
the change of behavior from the intermediate to the
SR regime takes place at $\sigma = 2-\eta_{SR}$.

Many other
studies also have considered  this problem of $\sigma=2$ with various
conclusions. In particular, van Enter \cite{enter} obtained that
for $n \geq 2$, for the classical and quantum
$XY$ models long--range perturbations are relevant in the
regime $\sigma=2$ in contradiction with Sak results. The same statement for arbitrary $n$ was made in Ref. \cite{gusmao}.
The Sak scenario was  also challenged in  Ref. \cite{pico}, which presents results of a Monte Carlo study for the ferromagnetic Ising model with long--range
interactions in two dimensions.  The author claims in addition that the results close to the change of regime from
intermediate to SR ($\sigma\geq 2$) do not agree with the renormalization group predictions.

A first numerical study of the exponent $\eta$ for $d = 2$
as a function of $\sigma$ has already been done in
Ref. \cite{luijten}. In particular, the authors obtained in the intermediate
regime ($d/2<\sigma<2$) a result well described by the exponent
$\eta = \eta_{G}=2-\sigma$ up to $2-\sigma = \eta_{SR}$ and $\eta = \eta_{SR}$ for
larger $\sigma$. In the subsequent paper  \cite{luijten2}, the authors claim that the boundary between
the SR and the LR fixed points for $d=2$ corresponds to $\sigma=7/4$.
In a field-theoretic approach \cite{nalimov} it was proved, to all orders in perturbation theory,
the stability of the SR fixed point for $\sigma>2-\eta_{SR}$
and of its LR counterpart for $\sigma<2-\eta_{LR}$, where
$\eta_{LR}$ is the anomalous dimension of the field, evaluated at
the long--range fixed point \cite{luijten2}.
Quite recent numerical analysis of the problem was presented in Ref. \cite{parisi}.
By including the subdominant power law, the numerical data are consistent with the standard renormalization
group (RG) prediction by Ref. \cite{sak}.

These debates have a practical importance for the case \eqref{3d}, corresponding to $\sigma=2$. If this case  is described by the SR  fixed point,
we have \cite{fisher} Eq. \eqref{si} and  (in quadratic expansion with respect to $\epsilon$)
\begin{eqnarray}
\frac{1}{\gamma_{SR}}=1-\left(\frac{n+2}{n+8}\right)\frac{\epsilon}{\sigma}-\frac{(n+2)(7n+20)}{(n+8)^3}{\cal Q}(\sigma)\left(\frac{\epsilon}{\sigma}\right)^2
\end{eqnarray}
with $\epsilon=2\sigma-d=1$, and
\begin{eqnarray}
{\cal Q}(\sigma)=\sigma\left[\psi(1)-2\psi\left(\frac{\sigma}{2}\right)+\psi(\sigma)\right],
\end{eqnarray}
where $\psi$ is the logarithmic derivative of the gamma function.
If this case  is described by the LR  fixed point,  we have the $\varepsilon$-expansion for the critical exponents of the  $O(n)$ model,
\cite{ma} (in the same approximation):
\begin{eqnarray}
\gamma_{LR}&=&1+\frac{n+2}{2(n+8)^2}\epsilon+\frac{n+2}{4(n+8)^3}(n^2+22n+52)\epsilon^2\nonumber\\
\eta_{LR}&=&\frac{n+2}{2(n+8)^2}\varepsilon^2.
\end{eqnarray}
with $\epsilon=4-d=1$.
High temperature expansions  may supply important argument in the debates.

The presence of quenched disorder can qualitatively change critical behavior of a magnetic system.
 A general argument \cite{harris}
shows that one should expect a new type of critical behavior for the random system, distinct from that of a pure one,
whenever the specific heat of the pure system diverges at the transition temperature.
This certainly happens for both cases \eqref{2d} and \eqref{3d}.

Fixed point $O(\epsilon)$ where $\epsilon =4-d$ for short--range interaction model  with weak quenched disorder for $n>1$ was found  by Harris and Lubensky \cite{harris}. Critical exponents calculated for this critical point are
\begin{eqnarray}
\eta_{SR}&=&\frac{n(5n-8)}{256(n-1)^2}\epsilon^2\\
\gamma_{SR}&=&1+\frac{3n}{16(n-1)}\epsilon.
\end{eqnarray}

For $n=1$ Khmelnitskii found another disordered fixed point of order $O(\sqrt{\varepsilon})$ \cite{khmel}.
Critical exponents calculated for this critical point  \cite{khmel,lubensky,grinstein} are ($\epsilon>0$)
\begin{eqnarray}
\label{k}
\eta_{SR}&=&-\frac{\epsilon}{106} +O\left(\epsilon^{3/2}\right)\\
\label{k2}
\gamma_{SR}&=&1+\frac{1}{2}\left(\frac{6\epsilon}{53}\right)^{1/2}+O(\epsilon).
\end{eqnarray}
Results for magnetic susceptibility  and heat capacity in the critical region at $d=4$  can be summed up as \cite{kenna}
\begin{eqnarray}
\chi(T)&\sim &t^{-1}\exp\left[\left(D\ln t^{-1}\right)^{1/2}\right]\left(\ln t^{-1}\right)^{\hat{\gamma}}\\
C(T)&\sim &\exp\left[-2\left(D\ln t^{-1}\right)^{1/2}\right]\left(\ln t^{-1}\right)^{\hat{\alpha}},
\end{eqnarray}
where $D=6/53$, and $\hat{\alpha}=1/2,\hat{\gamma} =0$ \cite{aharony,jug,ballesteros}, $\hat{\alpha}=1.24,\gamma=-.4$ \cite{shalaev,geldart}.

We are aware of a single paper where a model containing both disorder and long--range interaction was studied \cite{theumann}.
There critical properties  of a random Ising model with long--range isotropic
interactions \eqref{long} were analysed by using renormalisation group methods in an
expansion in $\epsilon=2\sigma-d$. For $\epsilon > 0$ the critical behaviour was
described by a stable fixed point $O(\sqrt{\epsilon})$. Like in  previous papers considering no--quenched disorder case, it was found that when
$\sigma=2-\eta_{SR}$ the system crosses over smoothly to SR behaviour. The  peculiarity is
that for the random fixed point $\eta_{SR}<0$ (see Eq. \eqref{k}).
Thus the
crossover to SR behaviour is analysed and takes place when $\sigma = 2 + \epsilon/106$,
$4-d>0$. Thus, according to this work, the case \eqref{3d}  with added weak quenched disorder  corresponds to the LR fixed point.

It is  unclear  whether fixed point obtained for weak  disorder potential scattering added to Hamiltonian \eqref{mi} is the only possible one for models with disorder.
 We believe that critical exponent obtained by high--temperature expansions in \cite{cheianov}
($\gamma=1.4$)
is an indication of the existence of a new fixed point existing for
the strong gas disorder model considered there. Hence additional
studies of high temperature expansions in that model (with different dimensions, different exchange decay laws etc.) would be of much interest.

The authors are grateful to D. E. Khmelnitskii for very illuminating discussions.

\end{document}